## The Morphology and Circuity of Walkable and Drivable Street Networks

Geoff Boeing
Northeastern University
g.boeing@northeastern.edu

### Abstract

Circuity, the ratio of network distances to straight-line distances, is an important measure of urban street network structure and transportation efficiency. Circuity results from a circulation network's configuration, planning, and underlying terrain. In turn, it impacts how humans use urban space for settlement and travel. Although past research has examined overall street network circuity, researchers have not studied the relative circuity of walkable versus drivable circulation networks. This study uses OpenStreetMap data to explore relative network circuity. We download walkable and drivable networks for 40 US cities using the OSMnx software, which we then use to simulate four million routes and analyze circuity to characterize network structure. We find that walking networks tend to allow for more direct routes than driving networks do in most cities: average driving circuity exceeds average walking circuity in all but four of the cities that exhibit statistically significant differences between network types. We discuss various reasons for this phenomenon, illustrated with case studies. Network circuity also varies substantially between different types of places. These findings underscore the value of using network-based distances and times rather than straight-line when studying urban travel and access. They also suggest the importance of differentiating between walkable and drivable circulation networks when modeling and characterizing urban street networks: although different modes' networks overlap in any given city, their relative structure and performance vary in most cities.

## Introduction

Street networks organize and structure human spatial dynamics and flows in a city. They underlie commutes, discretionary trips, and the location decisions of households and firms. Accordingly, substantial research has been conducted in recent years to better characterize the topological and geometric characteristics of urban street networks (*1–3*). Topological character describes the configuration of the network and includes measures of connectivity, centrality, and clustering. Geometric character describes the network's distances, areas, and densities. Both intermingle to define the network's structure, efficiency, and performance.

Street network analysis has become an important bridge between graph theory and urban morphology and planning (*4–9*). Recent studies have found that cities can be clustered and classified according to their network's structural characteristics (*10–14*). In other words, cities' circulation networks exhibit spatial signatures that can be quantified and operationalized to differentiate types of places. One branch of this research literature augments urban morphology studies with graph-theoretic topological analyses of street networks (*12, 13, 15–17*). Another branch considers the implications of network configuration and geometry on transportation and circulation (*18–22*).

Circuity, the ratio of network distances to straight-line distances, is an important measure of network structure and transportation efficiency (*23*). Interest in circuity is not new. In 1929, the modernist polemic Le Corbusier (*24*) wrote: "The circulation of traffic demands the straight line; it is the proper thing for the heart of a city. The curve is ruinous, difficult and dangerous; it is a paralyzing thing… The winding road is the Pack-Donkey's Way, the straight road is man's way." In particular, Corbusier argued that planners must eradicate walkable, self-organized streets and paths from traditional cities to enable the development of deliberate, rational, straight-line roads for cars. Similar plans were enacted throughout the twentieth century, including Robert Moses's "meat ax" carving up New York's neighborhoods to make room for new expressways (*25–27*). Over time, evolving street network pattern and design standards that impact circuity have been bureaucratized by the ITE, the Federal Housing Administration, and the Urban Land Institute (*28–30*).

Circuity results from a network's planning, configuration, and underlying terrain (*31*). In turn, it impacts how humans use urban space for settlement and travel. Levinson and El-Geneidy (*32, 33*) demonstrate that households tend to select residential locations that offer less circuitous work commutes. Giacomin and Levinson (*34*) show that average US metropolitan circuities rose between 1990 and 2000, as recent peripheral suburban development has featured more circuitous street network designs. O'Sullivan and Morrall (*35*) find that walking trip circuity (counterintuitively) increases in Calgary during winter months, as pedestrians avoid shortcuts that have been rendered impassible by ice. Most circuity research has treated a city's street network as a single entity. However, multiple circulation networks – that may be disambiguated by mode (e.g., walking, driving, biking) – overlap to constitute the city's complete multimodal circulation network.

Past research has measured the circuity of car (*34, 36*), bike (*37, 38*), and transit (*23*) networks. However, the relative circuity of walkable networks versus drivable networks has been underexplored. On one hand, car-only routes like freeways can provide straight paths to link opposite sides of the city by cutting across grids and winding surface streets. On the other hand, such routes may have circuitous elements such as cloverleaf interchanges and are engineered to optimize trip time – a function of distance and speed – rather than distance itself (cf. *39*). Drivable networks may also include one-way streets, but pedestrians can traverse them bidirectionally. Moreover, walking networks provide paths across parks, mid-block cut-throughs, passageways between buildings, and other shortcuts that driving networks lack. The relative circuity of drivable versus walkable networks depends on the magnitude of the effect that each's unique features have on enabling trips to approximate straight-line travel. This effect likely varies by city, as a function of topography, planning history, and circulation network design (*36, 40*).

Although past research has examined overall street network circuity in various cities (*23, 32–38*), researchers have not studied the relative circuity of walking versus driving networks across multiple cities. This study uses open-source software and open data to explore relative network circuity. We download OpenStreetMap data for walkable and drivable circulation networks in 40 US cities using the OSMnx software, which we then use to simulate four million routes and analyze each's circuity. We find statistically significant differences between driving circuity and walking circuity in all but two cities. Driving circuity exceeds that of walking in all but four of the cities with statistically significant differences. Moreover, circuity varies in different kinds of places. These findings suggest the importance of using network-based distances to study urban access as well as the importance of differentiating between circulation network types. Although walkable and drivable networks overlap in every city, their relative structure and performance differ in most cities.

The following section briefly introduces the mathematics of graph theory and the data and tools used to study street networks. Then we present the methods used to analyze relative circuity and the results of this analysis. Finally, we discuss these results in the context of urban morphology and planning.

## Analytical Background

Street network analysis uses the mathematics of graph theory (*41–43*). A graph $G = (V, E)$ is composed of a set $V$ of vertices connected to one another by a set $E$ of edges (*44*). An undirected graph's edges point mutually in both directions, but a directed graph's edges are one-way, from an origin vertex to a destination vertex. Multigraphs allow multiple edges between a pair of graph vertices. A planar graph can be represented in two dimensions with edges only intersecting at vertices – otherwise, it is nonplanar. Street networks are nonplanar graphs due to bridges and underpasses. Primal representations model street segments as edges and intersections and dead-ends as vertices (*11, 45*). This study models drivable street networks as primal, nonplanar, directed, multigraphs weighted by length, and walkable networks as the same but undirected. Street networks are spatial graphs. Their vertices and edges are embedded in space and, in turn, have geometric characteristics such as circuity that rely on lengths and areas – alongside the standard topological traits of all graphs (*46*).

Street network data traditionally come from various sources, including disparate municipal and state repositories, expensive commercial data sets, and (in the US) the census bureau's TIGER/Line roads shapefiles. Many studies rely on the latter because of its accessibility and comprehensiveness. However, TIGER/Line suffers from substantial spatial inaccuracies, broad classifiers that lump multiple path types together, and the misrepresentation of traffic-diverting bollards as through-streets (*47*). The latter in particular biases routing results. OpenStreetMap offers an alternative data source. It is an online collaborative mapping project that covers the entire world (*48, 49*). As of 2017, OpenStreetMap contained over 4.4 billion geospatial objects in its database, along with over 1.5 billion tags which describe these features. The objects comprise streets, trails, building footprints, land parcels, rivers, power lines, points of interest, and many other features.

These data are added to the OpenStreetMap database in typically one of two ways. The first is through large-scale imports of publicly available data sources, such as census TIGER/Line data or municipalities' shapefiles. The second is through the many individual additions and edits performed on an ongoing basis by OpenStreetMap's users and contributors. OpenStreetMap's data are largely high quality (*50–56*). In 2007, it imported the TIGER/Line roads as a foundation, and since then, the user community has added additional features, richer attribute data, and spatial corrections (*57*). Of particular relevance to the present study, OpenStreetMap data go far beyond those available in TIGER/Line as they include pedestrian paths, park trails, passages between buildings and through blocks, and finer-grained codes classifying various path and street types.

Researchers typically acquire street network data from OpenStreetMap in one of three ways. First, Overpass provides an API that allows users to query for geospatial features. However, its query language is somewhat difficult to use directly. Second (and accordingly), a handful of commercial services have sprung up as middle-men, downloading data extracts for certain areas or bounding boxes and then providing them to users. However, these services are often expensive, slow, and not customizable. While they may work well for studying the street network within a single bounding box, they are inconvenient for acquiring data in multiple precisely-bounded study sites. Further, they provide data as geometric shapefiles, which do not lend themselves immediately to nonplanar, graph-theoretic network analysis.

A third method for acquiring OpenStreetMap network data is OSMnx. OSMnx is a free, open-source Python package for downloading and analyzing street networks from OpenStreetMap (*58*). It can query by bounding box, address plus network distance, polygon (e.g., from a shapefile), or by place name (which resolves to a polygon representing the place's borders) such as cities, boroughs, or counties. OSMnx can download drivable, walkable, or bikeable networks, as well as other infrastructures such as power lines or subway systems.

Walkable and drivable paths are identified by OpenStreetMap metadata. Once the network has been downloaded, OSMnx automatically assembles it into a nonplanar directed multigraph and corrects its topology to retain vertices only at true intersections and dead-ends. This simplification process faithfully retains the true geometry and length of each street segment. Finally, OSMnx can analyze these street networks in various ways, including shortest-path calculations, topological measures such as centrality and clustering, and geometric measures such as intersection density and circuity (*16*).

## Methods

This study uses OSMnx to calculate shortest path distances along walkable and drivable street networks in various cities. It builds on the research designs of (*32–34*) by simulating 100,000 routes in each city and calculating circuities as a function of network distance and straight-line distance. Unlike some previous morphological studies, we use nonplanar networks as a superior representation of topology (*58, 59*) and we utilize the fine-grained classifier codes in OpenStreetMap data to create separate (directed) driving and (undirected) walking networks that are more detailed that those of most prior studies. To study the difference in circuity between walkable and drivable networks, we examine 40 US cities. We select cities across the breadth of the nation, including most of the largest cities as well as several medium cities, small towns, and suburbs for contrast. For each city, we draw a convex hull around the municipal borders, then download the street network within this hull.

This technique offers two advantages. First, it allows us to focus on municipalities – the scale of urban planning jurisdiction and decision making – and their immediate vicinities without including the suburbs, exurbs, and urban fringe at the periphery of the broader metropolitan area. Second, it adjusts for substantial quirks in the shapes of municipal borders. Some cities' borders snake along a narrow linear feature to connect two disparate hemispheres. Others exhibit concavity and bend around large elbow curves. These quirks would cause inflated circuity scores as trips are forced to route around city borders instead of taking shorter and more direct routes that briefly cross through a neighboring town. Convex hulls solve this problem while still constraining the analysis to a city and its immediate environs. They also help us mostly avoid large bodies of water around which some metropolitan areas wrap – such as the San Francisco Bay and the Puget Sound – and which significantly impact metropolitan-scale circuity.

To acquire the street network data, we use OSMnx to download each city's walkable and drivable street networks – constrained to the convex hull – from OpenStreetMap. OSMnx uses OpenStreetMap's fine-grained tags to identify walkable and drivable streets and paths. In case of a disconnected network, we retain only the largest connected component. Then, for each city and network type, we simulate 50,000 random routes. This number of simulations was arrived at after a sensitivity analysis revealed that the means typically converge at stable values around this number. These randomized origin-destination pairs need not reflect the spatial distribution and weighting of real-world trips, as our goal is instead to characterize the structure of each network as a whole rather than just its most well-worn paths.

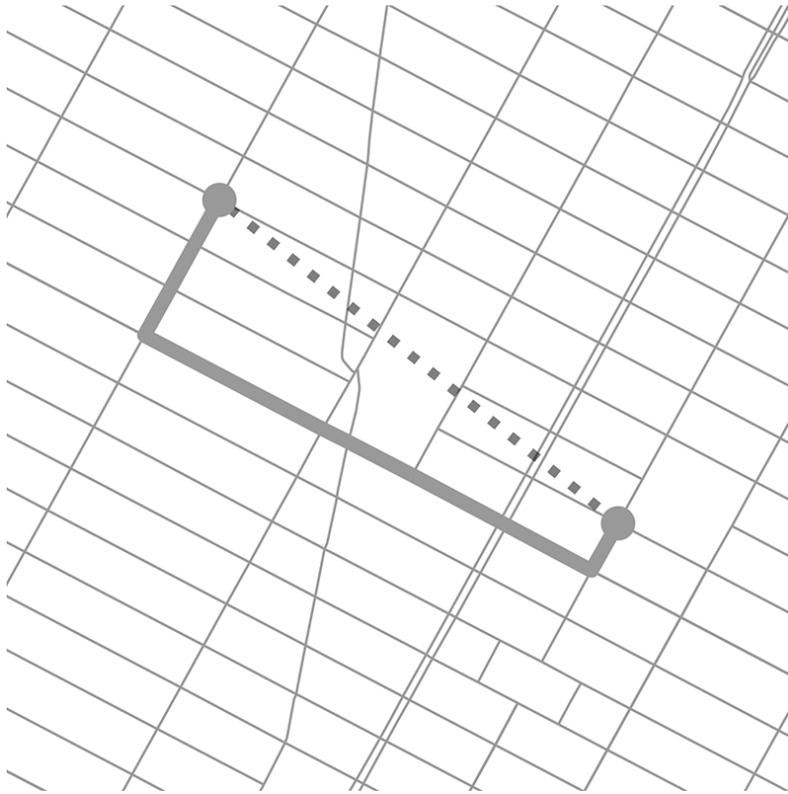

**Figure 1.** A section of Manhattan's drivable network, showing the shortest path (thick solid line) between two vertices, accounting for one-way streets, and the great-circle path (thick dashed line).

For each simulation, we randomly select two vertices, calculate the shortest path between them using Dijkstra's algorithm (*60*), and then calculate the great-circle distance between these two vertices. The great-circle distance $\zeta_{\mathrm{gc}}$ is calculated as:

$$\zeta_{\mathrm{gc}} = r \, \arccos(\sin(\Phi_1) \, \sin(\Phi_2) \, + \, \cos(\Phi_1) \, \cos(\Phi_2) \, \cos(|\lambda_1 - \lambda_2|))$$

where $r$ represents the Earth's radius of approximately 6371 km and $\Phi_1$, $\lambda_1$, $\Phi_2$, and $\lambda_2$ represent, in radians, the geographical latitude and longitude of two points. The great-circle distance $\zeta_{\mathrm{gc}}$ thus represents the shortest distance along the curved surface of the earth, and is more accurate than the Euclidean distance. We calculate each walking or driving route's circuity as:

$$\psi = \frac{\zeta_{\mathrm{net}}}{\zeta_{\mathrm{gc}}}$$

where, for each route, $\psi$ represents circuity, $\zeta_{\mathrm{net}}$ represents the shortest-path network distance between the origin and destination vertices, and $\zeta_{\mathrm{gc}}$ represents the great-circle distance between these vertices. Thus, a route's circuity is the ratio of the shortest-path network distance to the great-circle distance between the origin and destination. Figure 1 illustrates the difference between a shortest-path route and a straight line between two vertices in Manhattan's driving network, accounting for one-way streets.

We hypothesize that the average circuity of a city's walkable circulation network differs from that of its drivable circulation network. To test this hypothesis, for each city, we conduct a *t*-test to ascertain the statistical significance of the difference between the simulated walking routes' average circuity and that of the simulated driving routes, to see if we may reject the null hypothesis $H_0$:

$$H_0: \mu_w = \mu_d$$
$$H_1: \mu_w \neq \mu_d$$

where, for each city, $\mu_w$ is the mean $\psi$ of the routes along its walking network and $\mu_d$ is the mean $\psi$ of the routes along its driving network. We conduct two-sided *t*-tests because it is not known *a priori* if $\mu_w$ is expected to be greater than $\mu_d$ or vice-versa. For instance, if the effects of pedestrian pathways and cut-throughs exceed the effect of freeways on minimizing trip circuity, we would expect $\mu_d$ to be greater than $\mu_w$ (that is, walking allows for more direct routes). Conversely, if the effect of freeways exceeds drawbacks of one-way streets and the effects of pedestrian pathways and cut-throughs on minimizing trip circuity, we would expect $\mu_w$ to be greater than $\mu_d$ (that is, driving allows for more direct routes). Next, for each city, we calculate an indicator $\varphi$ to represent the ratio of average driving circuity to average walking circuity:

$$\varphi = \frac{(\mu_d - 1)}{(\mu_w - 1)}$$

We subtract 1 from each term in the $\varphi$ ratio because the minimum possible circuity is 1 (cf. *34*). The following section presents the results of these route simulations and the resulting statistical analyses.

## Results

Table 1 presents the average circuity of walking routes ($\mu_w$) and driving routes ($\mu_d$) in each city, with significance levels denoted by asterisks in the table. It also presents the mean distance of routes along the driving network ($\delta_d$), the mean distance of routes along the walking network ($\delta_w$), and the $\varphi$ ratio expressed as a percentage. We find that the average circuities of driving routes and walking routes differ by a statistically significant margin in 38 out of the 40 cities studied. Moreover, we find that the average driving circuity exceeds that of walking in 34 out of the 38 cities that have a statistically significant difference.

**Table 1**. Average Circuity of Walking and Driving Routes

| | $\mu_d$ | $\mu_w$ | $\delta_d$ (km) | $\delta_w$ (km) | $\varphi$ | |
|---|---|---|---|---|---|---|
| Atlanta, GA | 1.243 | 1.226 | 12.49 | 11.68 | 7.3% | *** |
| Baltimore, MD | 1.232 | 1.221 | 8.68 | 8.24 | 4.8% | *** |
| Boston, MA | 1.255 | 1.191 | 9.14 | 7.89 | 33.9% | *** |
| Charlotte, NC | 1.267 | 1.248 | 19.71 | 19.14 | 7.7% | *** |
| Chicago, IL | 1.194 | 1.178 | 18.79 | 18.40 | 8.8% | *** |
| Cincinnati, OH | 1.341 | 1.332 | 12.35 | 11.51 | 2.6% | *** |
| Cleveland, OH | 1.213 | 1.208 | 12.26 | 11.53 | 2.3% | *** |
| Dallas, TX | 1.177 | 1.180 | 24.23 | 24.33 | -1.5% | *** |
| Denver, CO | 1.242 | 1.209 | 15.19 | 13.23 | 15.3% | *** |
| Detroit, MI | 1.190 | 1.178 | 13.03 | 12.36 | 6.3% | *** |
| Fargo, ND | 1.336 | 1.291 | 7.64 | 7.00 | 15.6% | *** |
| Gary, IN | 1.324 | 1.285 | 7.90 | 6.93 | 13.6% | *** |
| Kansas City, MO | 1.223 | 1.339 | 21.51 | 23.08 | -34.2% | *** |
| Las Vegas, NV | 1.272 | 1.268 | 13.72 | 14.02 | 1.3% | ** |
| Los Angeles, CA | 1.198 | 1.196 | 28.47 | 27.93 | 0.6% | |
| Louisville, KY | 1.274 | 1.254 | 18.85 | 17.47 | 7.7% | *** |
| Manhattan, NY | 1.209 | 1.142 | 7.57 | 6.87 | 47.6% | *** |
| Miami, FL | 1.259 | 1.246 | 7.83 | 7.93 | 5.5% | *** |
| Minneapolis, MN | 1.241 | 1.223 | 8.24 | 7.40 | 8.4% | *** |
| Orlando, FL | 1.306 | 1.300 | 13.60 | 13.79 | 2.0% | *** |
| Philadelphia, PA | 1.200 | 1.184 | 12.85 | 12.79 | 8.9% | *** |
| Phoenix, AZ | 1.256 | 1.224 | 25.53 | 24.41 | 14.0% | *** |
| Pittsburgh, PA | 1.345 | 1.328 | 9.58 | 9.18 | 5.3% | *** |
| Portland, ME | 1.389 | 1.347 | 6.09 | 6.15 | 12.2% | *** |
| Portland, OR | 1.289 | 1.264 | 12.05 | 11.97 | 9.5% | *** |
| Redmond, WA | 1.522 | 1.396 | 5.79 | 5.18 | 31.8% | *** |
| Riverside, CA | 1.312 | 1.289 | 10.78 | 11.30 | 7.9% | *** |
| Salem, MA | 1.505 | 1.487 | 4.83 | 4.66 | 3.8% | *** |
| San Antonio, TX | 1.218 | 1.199 | 22.43 | 21.77 | 9.9% | *** |
| San Diego, CA | 1.307 | 1.354 | 25.90 | 25.91 | -13.3% | *** |
| San Francisco, CA | 1.308 | 1.210 | 11.46 | 6.96 | 46.2% | *** |
| Scranton, PA | 1.376 | 1.349 | 5.09 | 5.08 | 7.7% | *** |
| Seattle, WA | 1.289 | 1.251 | 12.29 | 11.08 | 15.1% | *** |
| St Augustine, FL | 1.373 | 1.331 | 4.39 | 3.88 | 12.9% | *** |
| St Louis, MO | 1.204 | 1.193 | 8.55 | 7.73 | 5.4% | *** |
| Stamford, CT | 1.340 | 1.340 | 6.93 | 6.63 | 0.1% | |
| Sugar Land, TX | 1.523 | 1.405 | 7.92 | 7.15 | 29.2% | *** |
| Tampa, FL | 1.281 | 1.267 | 14.85 | 13.99 | 5.3% | *** |
| Vicksburg, MS | 1.363 | 1.379 | 6.74 | 6.16 | -4.1% | *** |
| Walnut Creek, CA | 1.470 | 1.392 | 5.60 | 5.29 | 19.9% | *** |

NOTE: $\mu_w$ = mean circuity of the simulated routes along the walking network, $\mu_d$ = mean circuity of the simulated routes along the driving network, $\delta_d$ = mean distance (km) of routes along the driving network, $\delta_w$ = mean distance (km) of routes along the walking network, and $\varphi$ represents how much $\mu_d$ exceeds $\mu_w$ expressed as a percentage. Finally, ** indicates a statistically significant difference between $\mu_w$ and $\mu_d$ at the .01 level, and *** indicates significance at the .001 level.

The mean distance of routes along the driving network, $\delta_d$, and the mean distance of routes along the walking network, $\delta_w$, demonstrate how much distance can be saved, on an average trip, by taking the more direct mode of travel. For instance, in Manhattan, the average walking route is 0.7 kilometers shorter than the average driving route. However, the values of $\delta_w$ and $\delta_d$ are correlated with street network size, so cities with larger spatial extents demonstrate higher trip distances on average (cf. *61*).

To adjust for this scaling effect, $\mu_d$ and $\mu_w$ represent circuity as the mean of the simulated $\psi$ values (each route's ratio of shortest-path network distance to great-circle distance). Cities with orthogonal street grids or robust radial freeway systems tend to have the least circuitous average driving routes: in Philadelphia, Chicago, Detroit, Dallas, and Los Angeles the average driving route is only 18% to 20% more circuitous than the straight-line distance from origin to destination. More-recently developed suburbs with curvilinear residential street networks have the most circuitous average driving routes: in Sugar Land and Redmond the average driving route is over 52% more circuitous than the straight-line distance. Similar effects are seen in the walking networks. The average walking route in Manhattan, Chicago, and Detroit is less than 18% more circuitous than the straight-line distance from origin to destination. However, the average walking route in Sugar Land and Redmond are 40% more circuitous than the straight-line distance.

The primary focus of this study is on the relationship between driving network circuity and walking network circuity ($\varphi$ in Table 1). Manhattan and San Francisco exhibit the greatest values of $\varphi$. In each of these cities, the average driving route is over 46% more circuitous than the average walking route. In contrast, in San Diego and Kansas City, the average driving route is 13% and 34% less circuitous, respectively, than the average walking route.

## Discussion

These statistical results suggest that in most cities, the average circuity of the walking routes differs from that of the driving routes. In 38 of the 40 cities studied, this difference is statistically significant and the average circuity of driving routes exceeds that of walking routes in all but four of the cities with statistically significant results. In other words, on average, driving routes tend to be more circuitous than walking routes in most cities.

To interpret these network circuity findings, we use Manhattan and San Diego as illustrative cases. Manhattan's average driving route is 48% more circuitous than its average walking route. Figure 2 shows Manhattan's drivable street network on the left and its walkable street network on the right. We can immediately see that the walking network is much denser than the driving network. While the walking network contains 11,857 vertices and 1,331 kilometers of streets/paths (*n.b.* physical streets are equivalent to edges in an undirected graph), the driving network contains only 4,889 vertices and 1,064 kilometers of streets. Although it excludes expressways around the periphery of the island, the walking network provides numerous mid-block passages, pedestrian walkways, and frequent juncture points that in aggregate allow for more direct routes. In particular, Central Park poses an obstacle to straight-line driving, but the park's dense mesh of walking paths provides many cross-cutting and diagonal routes for pedestrians whose origins and destinations lie on either side of it. Manhattan also has many one-way streets that pedestrians may traverse bidirectionally, improving walking efficiency. In this case, the findings suggest that the effects of pedestrian pathways, cut-throughs, and bidirectionality exceed the effect of car-only motorways on trip directness.

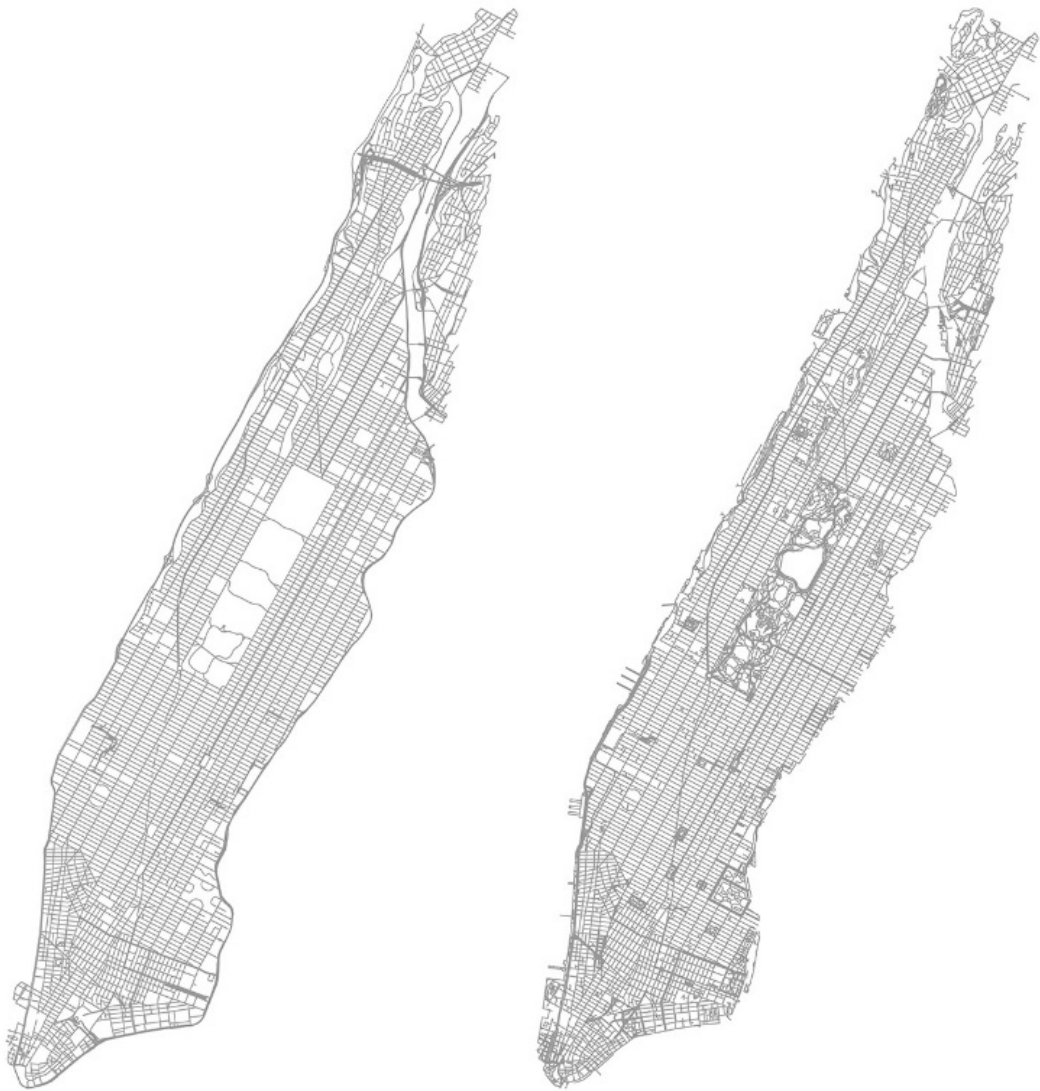

**Figure 2**. Manhattan's driving network (left) and walking network (right). Note that its convex hull includes a sliver of the Bronx.

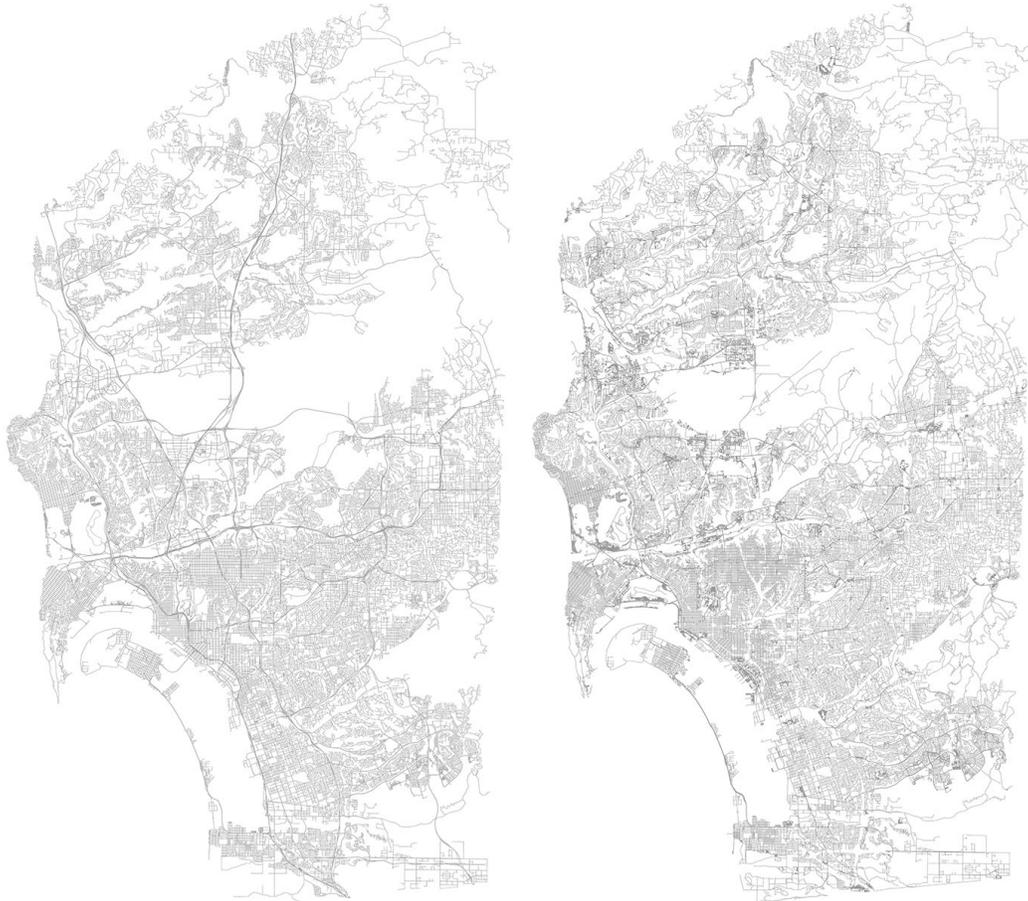

**Figure 3**. San Diego's driving network (left) and walking network (right).

As a contrasting example – unlike Manhattan and most of the cities studied – San Diego's $\varphi$ value is both negative and statistically significant. Its average driving route is 13% less circuitous than its average walking route. Figure 3 shows San Diego's drivable street network on the left and its walkable street network on the right. As in Manhattan, we can see that the walking network is denser than the driving network. While the walking network has 94,118 vertices and 14,293 kilometers of streets/paths, the driving network has only 51,544 vertices and 11,283 kilometers of streets.

There seem to be three primary reasons why San Diego has a negative $\varphi$ value. First, its highway system runs at cross-cutting diagonals, providing direct routes across the city that local streets cannot match. Second, its hills and canyons create substantial open space, natural preserves, and in turn network thresholds. The walking network provides dendritic, circuitous access into these spaces without offering efficient links across them. Third, Coronado Island (technically a peninsula connected to the mainland by a trombolo) connects to downtown San Diego via the San Diego–Coronado Bridge for reasonably direct driving routes. However, this bridge is automobile-only, so walking routes between Coronado and the majority of the city must route along the peninsula and around the entire San Diego Bay (similarly, the Kansas City network has 7 driving bridges over the Missouri River but only 1 walkable bridge). In other words, due to topography and design, San Diego relies particularly heavily on its freeway system for shortest-path connectivity. Accordingly, in this case, the findings suggest that the effect of car-only motorways exceeds the effects of pedestrian paths and passageways on route directness. Nevertheless, we find that this effect is the exception rather than the rule across our study sites.

Empirical studies in the urban economics, planning, and transportation literatures today often rely on Euclidean or great-circle distances to model accessibility. These findings add to the growing body of evidence that such straight-line measures are both inaccurate and inconsistent. Average network distances are at least 14% and sometimes over 50% longer than the straight-line distance from origin to destination. The magnitude of this phenomenon varies substantially between, for instance, Chicago and Sugar Land. This variation results from different topographies, planning eras, transportation technologies, and design paradigms. Thus, researchers and practitioners should use network-based distances and travel times to prevent biased distance measures in different kinds of places.

Moreover, we find that route circuity along walkable versus drivable networks differs significantly in most cities. A city's circulation network cannot be accurately modeled as a single monolithic entity. Rather, it comprises an interwoven set of overlapping circulation networks available to different modes of travel. Some one-way streets in the driving network may be accessible in both directions to pedestrians in the walking network. Similarly, pedestrian paths and mid-block passages may be unavailable to drivers, while expressways and certain bridges may be unavailable to pedestrians. Moreover, driving networks are more likely to be engineered to minimize travel time at high speeds rather than to minimize distance: distance itself is more important for modes with lower travel speeds. Measuring circuity depends on carefully defining what the circulation network includes and which modes of travel are of interest to the study. Thus, urban street network structure and performance cannot be sufficiently assessed without specifying network types and travel modes.

## Conclusion

This study examined the relative circuity of walkable and drivable urban circulation networks by simulating four million routes using OpenStreetMap data and the OSMnx software. It found that, in most cities, driving networks tend to produce more circuitous routes than walking networks do. Specifically, average driving circuity exceeds average walking circuity in all but four of the cities that have statistically significant differences. Old, dense cities like Manhattan and San Francisco saw the greatest effects, with average driving routes over 46% more circuitous than average walking routes. Network circuity also varies substantially between different types of places. These findings underscore the value of using network-based distances and times rather than straight-line when studying urban travel and access. They also suggest the importance of differentiating between walkable and drivable circulation networks: although these networks overlap, their relative structure and performance vary in most cities.

This study used simulated route distances as an indicator of circuity and network efficiency. Travel time is another important measure of efficiency. Future research can weight walking and driving routes by travel time to compare how simulated trip times vary by mode in different places as a function of network structure. Simulated trips could also be weighted by the likelihood of each being a real-world trip, based on travel survey data. This study randomized routes to examine overall network structure, but using actual trips would shed light on real-world travel behavior circuity and reduce bias from random sampling when neighborhood trips may be more common. Travel surveys could also provide information about average trip distances by mode. Beyond simple measures of network access, incorporating impedances based on grade, streetscape, traffic, and other data would provide a superior representation of routes. Moreover, it would be useful to explore how common planar models affect the results of route analysis (*62*). Finally, this study focused on US cities, but future work could use OSMnx and OpenStreetMap's worldwide data to compare networks in other countries to investigate the structure and performance of older European cities or rapidly growing African and Asian cities.